\newcommand{\be}{\begin{equation}}
\newcommand{\ee}{\end{equation}}
\newcommand{\bea}{\begin{eqnarray}}
\newcommand{\eea}{\end{eqnarray}}

\newcommand{\vecvar}[1]{\mbox{\boldmath$#1$}}
\newcommand{\ri}{{\rm i}}
\newcommand{\rd}{{\rm d}}
\newcommand{\re}{{\rm e}}
\newcommand{\ket}[1]{\left|#1\right\rangle}
\newcommand{\bra}[1]{\left\langle#1\right|}

\newcommand{\eref}[1]{Eq.~(\ref{#1})}
\newcommand{\Eref}[1]{Equation (\ref{#1})}
\newcommand{\fref}[1]{Fig.~\ref{#1}}
\newcommand{\Fref}[1]{Figure \ref{#1}}

\newcommand{\Ima}{\mathop{\mathrm{Im}}}
\newcommand{\Tr}{\mathop{\mathrm{Tr}}}

\documentclass{jpsj2}
\title{Transient Oscillation of Currents 
in Quantum Hall Effect of Bloch Electrons}

\author{
Manabu \textsc{Machida}$^{1}$
\thanks{Email address: machida@iis.u-tokyo.ac.jp}, 
Jun \textsc{Goryo}$^{2}$
\thanks{E-mail address: jungoryo@phys.aoyama.ac.jp} and 
Naomichi \textsc{Hatano}$^{1}$
\thanks{E-mail address: hatano@iis.u-tokyo.ac.jp}
}

\inst{$^{1}$Institute of Industrial Science, The University of Tokyo, 
Komaba, Meguro, Tokyo 153-8505 \\
$^{2}$Department of Physics and Mathematics, Aoyama Gakuin University, 
5-10-1 Fuchinobe, Sagamihara, Kanagawa 229-8558}

\abst{
We consider the quantum Hall effect of two-dimensional electrons 
with a periodic potential and 
study the time dependence of the Hall and longitudinal currents 
when the electric field is applied abruptly.  We find that 
the currents oscillate in time with very large frequencies 
because of quantum fluctuation and 
the oscillations eventually vanish, for their amplitudes 
decay as $1/t$.  
}

\kword{integer quantum Hall effect, Chern number, linear response theory}

\begin{document}
\maketitle

\section{Introduction \label{intro}} 

It is renowned that the Hall conductance in two dimensional electron 
systems under a strong magnetic field is quantized to an integer or 
a fraction multiplied by $e^2/h$ with very high accuracy\cite{Prange}.  
The relations between the conductance and topological numbers were 
discussed extensively
\cite{Thouless82a,Avron83a,Niu85a,Aoki86a,Ishikawa86a,Ishikawa87a,Imai90}, 
since the topological 
numbers take quantized values exactly. In the present paper, 
we discuss the integral quantization in noninteracting Bloch states.  
Thouless, Kohmoto, Nightingale and den Nijs (TKNN) showed using 
the Kubo formula that the quantized Hall conductivity is represented 
by the Chern number, which is a topological number defined on 
the two-dimensional torus (\textit{i.e.}, the magnetic Brillouin zone)
\cite{Thouless82a,Kohmoto85a}.  
The same result is also obtained from the 
adiabatic approximation\cite{Kohmoto89a,Hatsugai97a,Goryo07a}.  
It would be an intriguing issue, 
at least from a purely theoretical point of view, that how the 
topologically quantized conductivity is modified when we go beyond 
the Kubo formula or the adiabatic approach.

Interest in the TKNN theory was renewed recently in the field of 
ultra-cold atomic gases.  The TKNN Hamiltonian is mapped to the 
Hamiltonian of a cold atomic gas trapped by a rotating optical 
lattice.  Rotating Bose-Einstein condensates in a co-rotating 
optical lattice was indeed experimentally realized 
recently\cite{Tung06}, which fueled the interest in the TKNN theory.  
The atomic gas system 
does not contain any perturbative effects coming from impurities or 
long range Coulomb type interactions.  Hence, compared with 
the electron system in the solid states, the atomic gas system is 
clean and the theoretical results of the TKNN theory can be applied 
without taking into account the corrections from such perturbations.  
An alternative method of applying an effective magnetic field to 
a cold atomic gas is also proposed.\cite{Jaksch03a,Mueller04a}  
This method utilizes the internal degrees of freedom of cold atoms 
instead of the rotation of the system.  
The Hofstadter butterfly\cite{Hofstadter76a}, which 
has been observed in a two-dimensional superlattice structure 
in a semiconductor heterojunction\cite{Liu91a,Geisler04a,Geisler05a}, 
is predicted to be studied more easily using cold atomic gases.  

In this paper, we focus on the effect of a suddenly applied dc 
electric field on the integer quantum Hall effect of 
Bloch electrons.\cite{Machida06a}  
The results are readily applied to the cold atomic gas trapped 
by a rotating optical lattice.  
We calculate the resulting current with the Kubo formula
\cite{Nakano56a,Kubo56a,Kubo57a,Greenwood58a}.  
The linear response theory for an abruptly applied 
dc field was particularly investigated by Greenwood\cite{Greenwood58a}.  
We here follow Greenwood's formulation of the linear response theory.  

An interesting feature of our finding is an observation of fluctuation 
around the quantized conductivity, which is normally considered 
a very rigid quantity; 
we find that the Hall current has a time-dependent correction term to 
the Chern-number term in the TKNN theory.  
The Hall current $j_{x}$ and 
the longitudinal current $j_{y}$ oscillate 
in time with large frequencies because of quantum fluctuation, 
oscillation between different subbands.  
The oscillation eventually ceases and 
the time-dependent Hall current converges 
to the Chern-number term of the TKNN theory.  
The amplitude of the oscillation decays as $1/t$.  
In the previous paper\cite{Machida06a}, we already reported 
the existence of time-dependent correction terms.  
In the present paper, we present additional calculations particularly 
on the long-time behavior and on the time-dependent fields under an 
applied current.  

This paper is organized as follows.  
In \S\ref{hami}, we derive the currents in the $x$ and $y$ directions 
following the Greenwood linear response theory.  
We derive the same results as in our previous paper, but under 
a different gauge.  
We also mention the correspondence between electron gases in a 
magnetic field and rotating cold atomic gases.  
In \S\ref{longtime}, we show that the time-dependent oscillation 
of the currents decays as $1/t$ and eventually ceases, and 
the Hall current approaches to a certain value obtained from 
the TKNN theory.  
Finally we give conclusions.  
In Appendix, we calculate electric fields under an applied current 
instead of currents under an applied field.  
We show that the voltages have similar time dependence.

\section{Time Dependence of Currents\label{hami}} 

We consider noninteracting electrons in a periodic potential in 
the $x$-$y$ plane.  
A magnetic field $B$ is applied in the $z$ direction.  
At time $t=0$, we suddenly apply an electric field $\vecvar{E}(t)$ 
in the $y$ direction.  
We calculate the currents of this system with the Kubo formula.  
The Kubo formula for a step-function external field is also known as 
the Greenwood linear response theory\cite{Greenwood58a}.  

Using the Landau gauge, we write the Hamiltonian of the system as 
\be
\mathcal{H}=\mathcal{H}_0-eyE_y\theta(t),
\label{hami:hamiltonian}
\ee
where
\bea
\mathcal{H}_0
&=&
\frac{1}{2m_{\rm e}}\left[p_x^2+(p_y+eBx)^2\right]+V_{\rm p},
\\
V_{\rm p}
&=&
U_0\cos\left(\frac{2\pi x}{a}\right)+
U_0\sin\left(\frac{2\pi y}{b}\right).
\eea
In our previous paper\cite{Machida06a}, we treated the external field 
as a time-dependent vector potential.  
We here use the time-dependent scalar potential as in 
\eref{hami:hamiltonian}.  
We show below that the resulting formulae are the same.  

The Hamiltonian (\ref{hami:hamiltonian}) can also describe 
a rotating dilute cold atomic gas trapped in an optical lattice
\cite{Landau}.  
To see this correspondence, let us consider a cold atom with mass 
$m_{\rm a}$ confined in a harmonic potential.  
The periodic optical lattice which traps the cold atom rotates 
in the $z$ direction with angular momentum $\Omega$.  At $t=0$, 
the optical lattice is tilted along the $y$-axis or accelerated 
in the $y$ direction.  
The Hamiltonian of this system is written in the rotating frame as
\bea
\tilde{\mathcal{H}}
&=&
\frac{1}{2m_{\rm a}}\left(p_x^2+p_y^2\right)-\vecvar{\Omega}\cdot\vecvar{L}+
\frac{1}{2}m\Omega^2\left(x^2+y^2\right)+V_{\rm p}-yV_y\theta(t),
\label{hami:coldatom}
\eea
where $\vecvar{\Omega}={^t}(0,0,\Omega)$ and 
$\vecvar{L}=\vecvar{p}\times\vecvar{x}$.  
We note that the centrifugal force is canceled because 
the frequency of the harmonic trap is the same as the frequency 
of the rotation, and the interactions between atoms are neglected.  
\Eref{hami:coldatom} is also expressed as 
\be
\tilde{\mathcal{H}}
=
\frac{1}{2m_{\rm a}}
\left[\left(p_x-m\Omega y\right)^2+\left(p_y+m\Omega x\right)^2\right]+
V_{\rm p}-yV_y\theta(t),
\ee
By substituting $m_{\rm e}$, $eB/2m_{\rm e}$, and $eE_y$ for 
$m_{\rm a}$, $\Omega$, and $V_y$, respectively (See Table \ref{hami:table}), 
we have 
\be
\tilde{\mathcal{H}}=
\re^{\ri(eB/2\hbar)yx}\mathcal{H}\re^{-\ri(eB/2\hbar)yx}.
\ee
Thus, moving from the Landau gauge to the symmetric gauge by 
the operator $\exp[\ri(eB/2\hbar)yx]$, 
we see that the Hamiltonian (\ref{hami:hamiltonian}) for an electron gas 
is identical to the Hamiltonian (\ref{hami:coldatom}) for a cold atomic gas.  
\begin{table}[tb]
\caption{Correspondence between Hamiltonians 
(\ref{hami:hamiltonian}) and (\ref{hami:coldatom}).}
\label{hami:table}
\begin{tabular}{cc}
\hline
Electron gas in a magnetic field & 
Rotating cold atomic gas \\
\hline
$m_{\rm e}$ & $m_{\rm a}$ \\
\hline
$eB/2m_{\rm e}$ & $\Omega$ \\
\hline
$eE_y$ & $V_y$ \\
\hline
\end{tabular}
\end{table}

We consider the ratio $\phi\,(=\Phi/\Phi_0)$ of 
the flux $\Phi\,(=Bab)$ per unit cell 
to the flux quanta $\Phi_0\,(=h/e)$.  We put 
\be
\phi=\frac{p}{q},
\ee
where $p$ and $q$ are coprime integers.  
Because of the presence of the periodic potential, 
each Landau level splits into $p$ sublevels.  

Let us first consider $\mathcal{H}_0$.  
We write the eigenvalues and eigenfunctions of $\mathcal{H}_0$ as 
\be
\mathcal{H}_0\ket{\phi_{Nm}}=E_{Nm}\ket{\phi_{Nm}},
\ee
where the subscript $N$ labels Landau levels and 
the subscript $m$ labels sublevels in a 
Landau level ($1\le m\le p$).  
We define the generalized crystal momentum $\hbar\vecvar{k}$ 
in the magnetic Brillouin zone:\cite{Thouless82a} 
$0\le k_x<2\pi/qa$ and $0\le k_y<2\pi/b$.  
Note that $\re^{\ri k_xqa}$ and $\re^{\ri k_yb}$ are the eigenvalues of 
the translational operator.  
We define 
\bea
\mathcal{H}_{0\vecvar{k}}
&\equiv& 
\re^{-\ri\vecvar{k}\cdot\vecvar{x}}\mathcal{H}_0
\re^{\ri\vecvar{k}\cdot\vecvar{x}},
\nonumber \\
\ket{\phi_{Nm}}
&\equiv& 
\re^{\ri\vecvar{k}\cdot\vecvar{x}}\ket{u_{Nm}(\vecvar{k})},
\eea
which satisfy 
\be
\mathcal{H}_{0\vecvar{k}}\ket{u_{Nm}(\vecvar{k})}=
E_{Nm}(\vecvar{k})\ket{u_{Nm}(\vecvar{k})}.
\ee
We thus block-diagonalized the Hamiltonian $\mathcal{H}_0$ into 
each subspace of $\vecvar{k}$.  

Let us consider small $U_0$ and treat 
the periodic potential as a perturbation in the subspace of 
a crystal momentum.  
Taking the lowest order terms into account, 
we obtain the wave function as\cite{Thouless82a}
\bea
u_{Nm}(\vecvar{k};x,y)
&=&
\sum_{n=0}^{p-1}d_{m}^n\sum_{s=-\infty}^{\infty}
\chi_N\left(x-qas-\frac{qan}{p}+k_y\ell^2\right)
\nonumber \\
&\times&
\re^{-\ri k_x(x-qas-qan/p)}\re^{-2\pi\ri (sp+n)y/b},
\label{hami:tknnwave}
\eea
where $\ell=\sqrt{\hbar/eB}$ is the cyclotron radius and 
$\chi_N(x)$ satisfies
\be
\partial_x^2\chi_N(x)=\left[
\frac{x^2}{\ell^4}-\frac{2N+1}{\ell^2}\right]\chi_N(x).
\ee
We note that $u_{Nm}(\vecvar{k};x,y)$ in \eref{hami:tknnwave} 
satisfies the magnetic Bloch theorem: 
\be
u_{Nm}(\vecvar{k};x+qa,y)\re^{2\pi\ri py/b}=
u_{Nm}(\vecvar{k};x,y)=
u_{Nm}(\vecvar{k};x,y+b).
\ee
We have the eigenenergy within the perturbation as 
\be
E_{Nm}(\vecvar{k})=
\hbar\omega_{\rm c}\left(N+\frac{1}{2}\right)+
\epsilon_{m}(\vecvar{k}),
\ee
where $\omega_{\rm c}$ is the cyclotron frequency.  
Here, $\epsilon_{m}(\vecvar{k})$ and $d_{m}^n$ satisfy the 
following secular equation (the Harper equation):
\cite{Thouless82a,Kohmoto89a}  
\be
U_0\re^{-\frac{\pi qb}{2pa}}
\cos\left(\frac{2\pi q}{p}n-\frac{qbk_y}{p}\right)
d_m^n
+\frac{U_0}{2}\re^{-\frac{\pi qa}{2pb}}
\left[d_m^{n+1}\re^{\ri k_xqa/p}+
d_m^{n-1}\re^{-\ri k_xqa/p}\right]=
\epsilon_{m}(\vecvar{k})d_m^n.
\label{hami:secular}
\ee
The coefficients satisfy $d_m^{n+p}=d_m^n$ and 
each Landau level splits into $p$ subbands.  

We consider the currents caused by the electric field $E_y\theta(t)$.  
In cold atomic gases, we can apply an effective electric field 
corresponding to $E_y\theta(t)$ either by making use of the 
gravitational force tilting the harmonic potential\cite{Anderson98a} 
or by accelerating the optical lattice\cite{Madison97a}.  
We calculate the currents in the $\alpha\,(=x,y)$ direction 
in the form
\be
j_{\alpha}(t)=
\mathop{\mathrm{Tr}}\rho(t)\frac{ev_{\alpha}}{qab},
\label{hami:currentdef}
\ee
where
\bea
v_x
&=&
\frac{1}{m_{\rm e}}p_x,
\nonumber \\
v_y
&=&
\frac{1}{m_{\rm e}}\left(p_y+eBx\right).
\eea
Here $\rho(t)$ is the density operator.  

Following Greenwood\cite{Greenwood58a}, we expand $\rho(t)$ with 
respect to the electric field $E_y$ and take the zeroth- and first-order 
terms into account: 
\be
\rho(t)\simeq\rho_0+\rho_1(t).
\ee
The zeroth term $\rho_0$ is the initial density operator, 
$\re^{-\beta\mathcal{H}_0}/\Tr\re^{-\beta\mathcal{H}_0}$.  
With the help of the von Neumann equation for the density operator, 
$\rho_1$ is calculated as 
\bea
\rho_1(t)
&=&
\frac{1}{\ri\hbar}\int_{-\infty}^t\rd t'\,
\re^{\ri\mathcal{H}_0(t'-t)/\hbar}
\left[-eyE_y\theta(t'),\rho_0\right]
\re^{-\ri\mathcal{H}_0(t'-t)/\hbar}
\nonumber \\
&=&
\frac{eE_y}{\ri\hbar}\int_0^t\rd t'\,
\re^{\ri\mathcal{H}_0(t'-t)/\hbar}
\left[-y,\rho_0\right]
\re^{-\ri\mathcal{H}_0(t'-t)/\hbar}.
\label{hami:rho1}
\eea
We note that the lower bound of the integral on 
the second line of \eref{hami:rho1} is zero 
because of the step function in the perturbation, whereas 
the lower bound is negative infinity 
in the TKNN theory\cite{Thouless82a}.  
By taking the trace in \eref{hami:currentdef} with respect to 
the states in \eref{hami:tknnwave}, we obtain the currents as 
\bea
j_{\alpha}(t)
&=&
\mathop{\mathrm{Tr}}\rho_1(t)\frac{ev_{\alpha}}{qab}
\nonumber \\
&=&
\ri\hbar
\frac{E_ye^2}{qab}\sum_{Nm}\sum_{N'm'}
\int_{0}^{2\pi/qa}\frac{\rd k_x}{2\pi/qa}
\int_{0}^{2\pi/b}\frac{\rd k_y}{2\pi/b}
\frac{f_{\rm F}\left(E_{Nm}(\vecvar{k})\right)}
{\left(E_{Nm}(\vecvar{k})-E_{N'm'}(\vecvar{k})\right)^2}
\nonumber \\
&\times&
\left\{
\bra{u_{Nm}(\vecvar{k})}v_{y\vecvar{k}}\ket{u_{N'm'}(\vecvar{k})}
\bra{u_{N'm'}(\vecvar{k})}v_{\alpha\vecvar{k}}\ket{u_{Nm}(\vecvar{k})}
\left[1-
\re^{-\ri\left(E_{Nm}(\vecvar{k})-E_{N'm'}(\vecvar{k})\right)t/\hbar}
\right]
-\mbox{c.c.}
\right\}
\nonumber \\
&=&
-E_y\frac{e^2}{2\pi\hbar}\sum_{Nm}\sum_{N'm'}\int_{\rm MBZ}
\frac{\rd\vecvar{k}}{2\pi}
\left[f_{\rm F}\left(E_{Nm}(\vecvar{k})\right)-
f_{\rm F}\left(E_{N'm'}(\vecvar{k})\right)\right]
\nonumber \\
&\times&
\mathop{\mathrm{Im}}
\biggl\langle
\frac{\partial u_{Nm}(\vecvar{k})}{\partial k_y}
\biggm|u_{N'm'}(\vecvar{k})
\biggr\rangle
\biggl\langle
u_{N'm'}(\vecvar{k})\biggm|
\frac{\partial u_{Nm}(\vecvar{k})}{\partial k_{\alpha}}
\biggr\rangle
\nonumber \\
&\times&
\left[1-
\re^{-\ri\left(E_{Nm}(\vecvar{k})-E_{N'm'}(\vecvar{k})\right)t/\hbar}
\right],
\eea
where MBZ stands for the magnetic Brillouin zone.  
Here we used 
$\mathcal{H}_{\vecvar{k}}=
\re^{-\ri\vecvar{k}\cdot\vecvar{x}}\mathcal{H}
\re^{\ri\vecvar{k}\cdot\vecvar{x}}$ 
and
\be
v_{\alpha\vecvar{k}}=
\re^{-\ri\vecvar{k}\cdot\vecvar{x}}v_{\alpha}
\re^{\ri\vecvar{k}\cdot\vecvar{x}}=
\frac{1}{\hbar}
\frac{\partial\mathcal{H}_{\vecvar{k}}}{\partial k_{\alpha}}.
\ee
Furthermore, noting the relation $v_y=[y,\mathcal{H}_0]$, we used 
\be
\bra{u_{Nm}(\vecvar{k})}\re^{-\ri\vecvar{k}\cdot\vecvar{x}}y
\re^{\ri\vecvar{k}\cdot\vecvar{x}}\ket{u_{N'm'}(\vecvar{k})}=
\frac{\ri\hbar}{E_{N'm'}(\vecvar{k})-E_{Nm}(\vecvar{k})}
\bra{u_{Nm}(\vecvar{k})}v_{y\vecvar{k}}\ket{u_{N'm'}(\vecvar{k})}.
\ee

Let us put the Fermi energy in a finite gap between 
the $m_0$th and $(m_0+1)$st subbands which belong to the lowest 
Landau level ($N=0$).  We consider the zero temperature.  Hence 
the Fermi distribution satisfies 
$f_{\rm F}(E_{Nm})=1$ if $N=0$ and $m\le m_0$, and 
$f_{\rm F}(E_{Nm})=0$ otherwise.  
Thus we obtain
\bea
j_{\alpha}(t)
&=&
-2E_y\frac{e^2}{2\pi\hbar}\int_{\rm MBZ}
\frac{\rd\vecvar{k}}{2\pi}\mathop{\mathrm{Im}}
\Biggl\{
\nonumber \\
&&
\sum_{m\le m_0}\sum_{m'>m_0}
\biggl\langle
\frac{\partial u_{0m}(\vecvar{k})}{\partial k_y}
\biggm|u_{0m'}(\vecvar{k})
\biggr\rangle
\biggl\langle
u_{0m'}(\vecvar{k})\biggm|
\frac{\partial u_{0m}(\vecvar{k})}{\partial k_{\alpha}}
\biggr\rangle
\nonumber \\
&\times&
\left[1-
\re^{-\ri\left(E_{0m}(\vecvar{k})-E_{0m'}(\vecvar{k})\right)t/\hbar}
\right]
\nonumber \\
&+&
\sum_{m\le m_0}\sum_{N'\ge 1,\,m'}
\biggl\langle
\frac{\partial u_{0m}(\vecvar{k})}{\partial k_y}
\biggm|u_{N'm'}(\vecvar{k})
\biggr\rangle
\biggl\langle
u_{N'm'}(\vecvar{k})\biggm|
\frac{\partial u_{0m}(\vecvar{k})}{\partial k_{\alpha}}
\biggr\rangle
\nonumber \\
&\times&
\left[1-
\re^{-\ri\left(E_{0m}(\vecvar{k})-E_{N'm'}(\vecvar{k})\right)t/\hbar}
\right]
\Biggr\}
\nonumber \\
&=&
-2E_y\frac{e^2}{2\pi\hbar}\int_{\rm MBZ}
\frac{\rd\vecvar{k}}{2\pi}\mathop{\mathrm{Im}}
\Biggl\{
\nonumber \\
&&
\sum_{m\le m_0}\sum_{m'>m_0}
\biggl\langle
\frac{\partial u_{0m}(\vecvar{k})}{\partial k_y}
\biggm|u_{0m'}(\vecvar{k})
\biggr\rangle
\biggl\langle
u_{0m'}(\vecvar{k})\biggm|
\frac{\partial u_{0m}(\vecvar{k})}{\partial k_{\alpha}}
\biggr\rangle
\nonumber \\
&\times&
\left[1-
\re^{-\ri\left(E_{0m}(\vecvar{k})-E_{0m'}(\vecvar{k})\right)t/\hbar}
\right]
\nonumber \\
&+&
\frac{\ell^2}{2}\sum_{m\le m_0}
\left(\ri\delta_{\alpha x}+\delta_{\alpha y}\right)
\left[1-
\re^{-\ri\left(E_{0m}(\vecvar{k})-E_{1m}(\vecvar{k})\right)t/\hbar}
\right]
\Biggr\},
\eea
where we used the fact that for $N'\ge 1$, we have
\be
\biggl\langle
\frac{\partial u_{0m}(\vecvar{k})}{\partial k_y}
\biggm|u_{N'm'}(\vecvar{k})
\biggr\rangle
\biggl\langle
u_{N'm'}(\vecvar{k})\biggm|
\frac{\partial u_{0m}(\vecvar{k})}{\partial k_{\alpha}}
\biggr\rangle=
\frac{\ell^2}{2}
\left(\ri\delta_{\alpha x}+\delta_{\alpha y}\right)
\delta_{N'1}\delta_{mm'}.
\ee
We can see that the time dependence of the current is due to the 
quantum fluctuations or quantum oscillations between various 
sets of discrete levels.  
We ignore the quantum fluctuation between $E_{0m}(\vecvar{k})$ and 
$E_{1m}(\vecvar{k})$ because its frequency, which is proportional 
to $E_{0m}(\vecvar{k})-E_{1m}(\vecvar{k})$, is very large compared to 
the frequency of the fluctuation between different subbands of the 
lowest Landau level, which is proportional to 
$E_{0m}(\vecvar{k})-E_{0m'}(\vecvar{k})$.  
Thus we obtain 
\bea
j_x(t)
&=&
\frac{E_ye^2}{2\pi\hbar}\left[N_{\rm Ch}+\Delta\sigma_x(t)\right],
\label{hami:jx} 
\\
j_y(t)
&=&
\frac{E_ye^2}{2\pi\hbar}\Delta\sigma_y(t),
\label{hami:jy}
\eea
where
\bea
N_{\rm Ch}
&=&
\sum_{m\le m_0}\int_{\rm MBZ}\frac{\rd^2\vecvar{k}}{2\pi\ri}
\left(
\biggl\langle
\frac{\partial u_{0m}(\vecvar{k})}{\partial k_x}
\biggm|
\frac{\partial u_{0m}(\vecvar{k})}{\partial k_y}
\biggr\rangle
-{\rm c.c.}\right),
\label{hami:nch}
\\
\Delta\sigma_x(t)
&=&
\sum_{m\le m_0}\sum_{m'> m_0}
\int_{\rm MBZ}\frac{\rd^2\vecvar{k}}{\pi}
\nonumber \\
&\times&
\Ima
\biggl\langle
\frac{\partial u_{0m}(\vecvar{k})}{\partial k_y}
\biggm|u_{0m'}(\vecvar{k})
\biggr\rangle
\biggl\langle
u_{0m'}(\vecvar{k})\biggm|
\frac{\partial u_{0m}(\vecvar{k})}{\partial k_x}
\biggr\rangle
\re^{\ri\left(\epsilon_{m'}(\vecvar{k})-\epsilon_{m}(\vecvar{k})\right)t/\hbar},
\label{hami:dsx}
\\
\Delta\sigma_y(t)
&=&
\sum_{m\le m_0}\sum_{m'> m_0}\int_{\rm MBZ}
\frac{\rd^2\vecvar{k}}{\pi}
\left|
\biggl\langle
\frac{\partial u_{0m}(\vecvar{k})}{\partial k_y}
\biggm|u_{0m'}(\vecvar{k})
\biggr\rangle
\right|^2
\sin\left[\left(
\epsilon_{m'}(\vecvar{k})-\epsilon_{m}(\vecvar{k})
\right)t/\hbar\right].
\label{hami:dsy}
\nonumber\\
\eea
Note that $N_{\rm Ch}$ is the Chern number and 
takes integer values.\cite{Thouless82a,Kohmoto85a}  
The time-dependent correction terms 
$\Delta\sigma_x(t)$ and $\Delta\sigma_y(t)$ 
express quantum fluctuation between different subbands of 
the lowest Landau level.  
These are expressed as the sum of different oscillating modes 
whose frequencies are determined by the energy difference 
$\epsilon_{m'}(\vecvar{k})-\epsilon_m(\vecvar{k})$.  

Hereafter, we show results of numerical calculation of 
the currents $j_x(t)$ and $j_y(t)$.  
Numerical calculation is carried out in a way similar to 
the Kubo formula for a dc field:\cite{Koshino04a,Koshino06a} 
the integrals in Eqs.~(\ref{hami:nch}), 
(\ref{hami:dsx}), and (\ref{hami:dsy}) are performed 
with random sampling of $k_x$ and $k_y$.  
In the calculation, we set $a=b$.  We here consider, for example, 
the following three cases: (i) $p/q=5/4$ and $m_0=2$ ($N_{\rm Ch}=2$), 
(ii) $p/q=7/6$ and $m_0=3$ ($N_{\rm Ch}=3$), and 
(iii) $p/q=7/6$ and $m_0=1$ ($N_{\rm Ch}=1$).

\begin{figure}[tb]
\begin{center}
\includegraphics[scale=1.0]{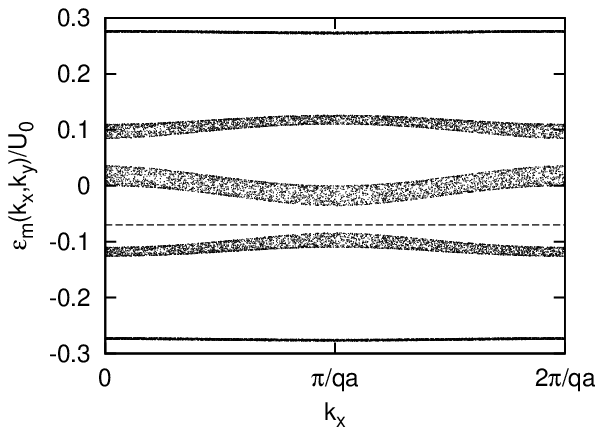} 
\includegraphics[scale=1.0]{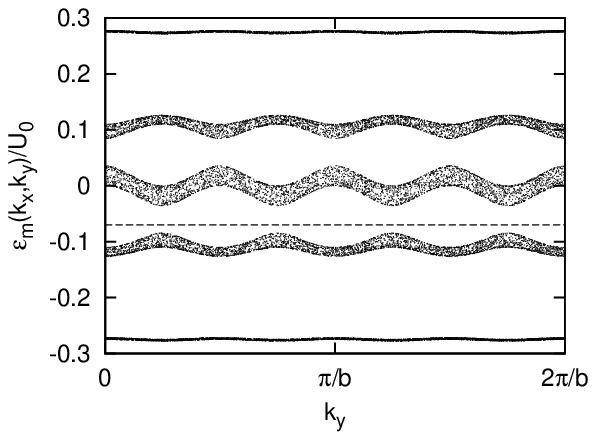} 
\end{center}
\caption{
Case (i): The band structure of $\epsilon_m$ as functions of 
$k_x$ (left) and $k_y$ (right).  The flux ratio $p/q=5/4$.  
We place the Fermi energy (the dashed line) between the second and 
third subbands.  
}
\label{hami:fig02a}
\end{figure}

\begin{figure}[tb]
\begin{center}
\includegraphics[scale=1.0]{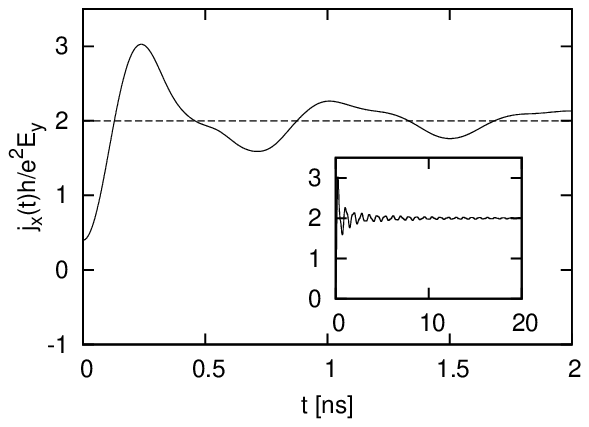}
\includegraphics[scale=1.0]{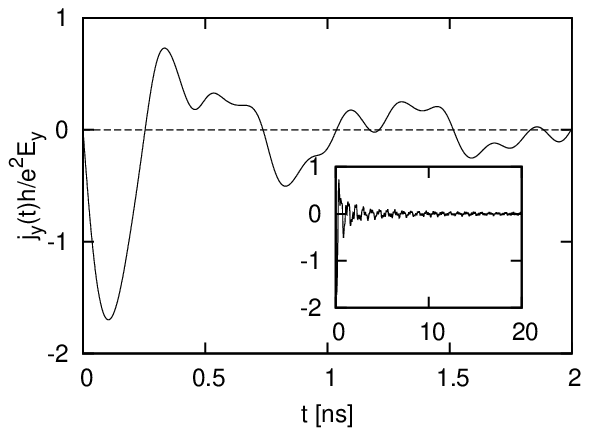}
\end{center}
\caption{
Case (i): The currents $j_x(t)h/e^2E_y$ and $j_y(t)h/e^2E_y$ 
are shown as functions of time.  
We set $p/q=5/4$, $m_0=2$, $U_0=0.1{\rm meV}$, and $a=b=100{\rm nm}$.  
The dashed lines show the convergent values of 
$j_x(t)h/e^2E_y$ and $j_y(t)h/e^2E_y$ 
($N_{\rm Ch}\,(=2)$ and $0$, respectively).  
The insets show long-time behaviors of 
$j_x(t)h/e^2E_y$ and $j_y(t)h/e^2E_y$.  
}
\label{hami:fig02b}
\end{figure}

The band structure in the case (i) is shown in \fref{hami:fig02a}.  
In the figure, the Fermi energy that we choose is plotted 
with the dashed line.  
\Fref{hami:fig02b} shows the currents $j_x(t)h/e^2E_y$ and 
$j_y(t)h/e^2E_y$ in the case (i).  
In the calculation, we put $U_0=0.1{\rm meV}$ and $a=b=100{\rm nm}$ 
as tipical values for 
quantum Hall systems on a semiconductor heterojunction.  
The currents oscillate irregularly reflecting the fact that 
the energy spectra $\epsilon_{m=2}(\vecvar{k})$ and 
$\epsilon_{m=3}(\vecvar{k})$ in \fref{hami:fig02a} strongly depend on 
$\vecvar{k}$, and 
$\Delta\sigma_x(t)$ and $\Delta\sigma_y(t)$  
are written as the sum of sinusoidal functions with 
different frequencies (Eqs.~(\ref{hami:dsx}) and (\ref{hami:dsy})).  
The insets show the long-time behavior of the currents.  
As we show in the next section, $\Delta\sigma_x(t)$ and 
$\Delta\sigma_y(t)$ vanish for large $t$.  

\begin{figure}[tb]
\begin{center}
\includegraphics[scale=1.0]{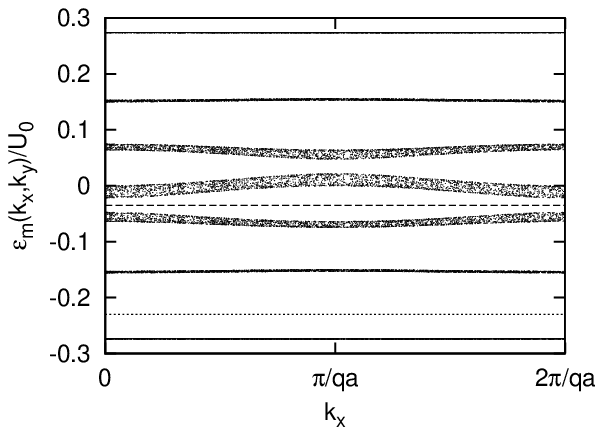}
\includegraphics[scale=1.0]{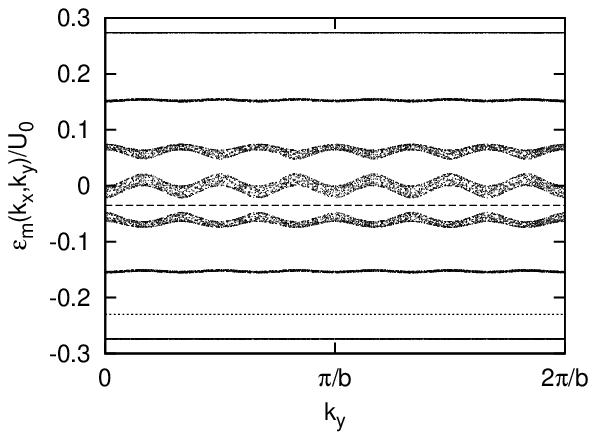}
\end{center}
\caption{
Cases (ii) and (iii): 
The band structure of $\epsilon_m$ as functions of 
$k_x$ (left) and $k_y$ (right).  The flux ratio $p/q=7/6$.  
We place the Fermi energy for the case (ii) between the third and 
forth subbands (the dashed line), and 
the Fermi energy for the case (iii) between the first and 
second subbands (the dotted line).  
}
\label{hami:fig03a}
\end{figure}

\begin{figure}[tb]
\begin{center}
\includegraphics[scale=1.0]{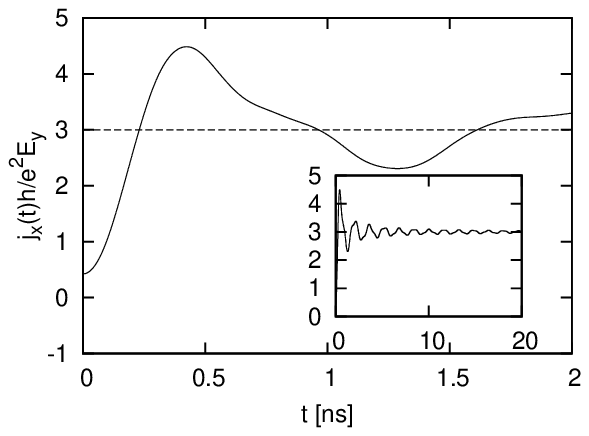}
\includegraphics[scale=1.0]{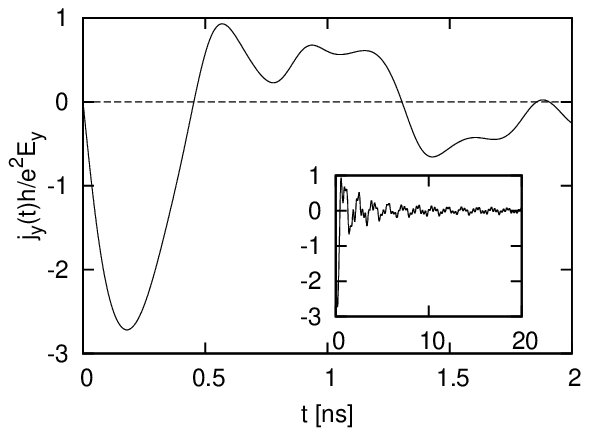}
\end{center}
\caption{
Case (ii): The currents $j_x(t)h/e^2E_y$ and $j_y(t)h/e^2E_y$ 
are shown as functions of time.  
We set $p/q=7/6$, $m_0=3$, $U_0=0.1{\rm meV}$, and $a=b=100{\rm nm}$.  
The dashed lines show the convergent values of 
$j_x(t)h/e^2E_y$ and $j_y(t)h/e^2E_y$ 
($N_{\rm Ch}\,(=3)$ and $0$, respectively).  
The insets show long-time behaviors of 
$j_x(t)h/e^2E_y$ and $j_y(t)h/e^2E_y$.  
}
\label{hami:fig03b}
\end{figure}

\begin{figure}[tb]
\begin{center}
\includegraphics[scale=1.0]{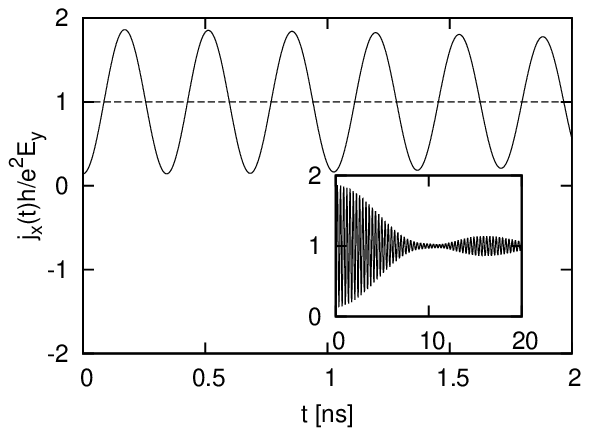}
\includegraphics[scale=1.0]{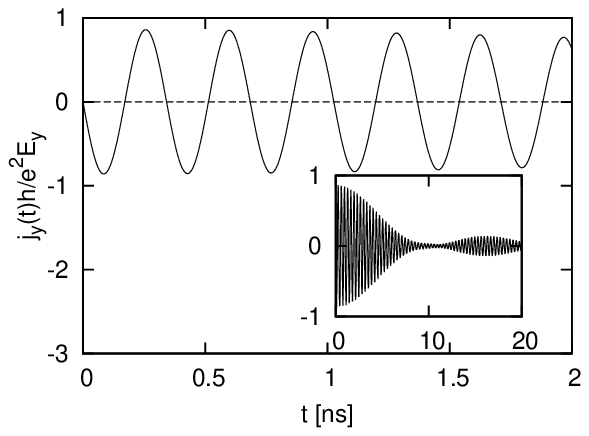}
\end{center}
\caption{
Case (iii): The currents $j_x(t)h/e^2E_y$ and $j_y(t)h/e^2E_y$ 
are shown as functions of time.  
We set $p/q=7/6$, $m_0=1$, $U_0=0.1{\rm meV}$, and $a=b=100{\rm nm}$.  
The dashed lines show the convergent values of 
$j_x(t)h/e^2E_y$ and $j_y(t)h/e^2E_y$ 
($N_{\rm Ch}\,(=1)$ and $0$, respectively).  
The insets show long-time behaviors of 
$j_x(t)h/e^2E_y$ and $j_y(t)h/e^2E_y$.  
}
\label{hami:fig05b}
\end{figure}

Similarly,  the band structure in the cases (ii) and (iii) is shown in 
\fref{hami:fig03a}.  
Since $N_{\rm Ch}$ in \eref{hami:nch} depends on $m_0$, $N_{\rm Ch}$ 
changes when we change the Fermi energy.  
In the figure, the Fermi energy for the case (ii) is plotted with 
the dashed line and that for the case (iii) is plotted with the 
dotted line.  
\Fref{hami:fig03b} shows the currents $j_x(t)h/e^2E_y$ and 
$j_y(t)h/e^2E_y$ in the case (ii).  The currents oscillate irregularly 
because of contributions from different frequencies.  
The insets show the long-time behavior of the currents.  
\Fref{hami:fig05b} shows the currents $j_x(t)h/e^2E_y$ and 
$j_y(t)h/e^2E_y$ in the case (iii).  In this case, the currents 
oscillate rather regularly 
because the first and second bands in \fref{hami:fig03a} are 
almost flat, and $\Delta\sigma_x(t)$ and $\Delta\sigma_y(t)$ 
are almost monochromatic.  
The insets show the long-time behavior of the currents.  

We remark the following three points.  
Firstly, the currents $j_x(t)$ and $j_y(t)$ are gauge invariant.  
We can also obtain the same results by using 
the time-dependent vector potential 
as we did in the previous paper\cite{Machida06a}.  
Secondly, if a dc current instead of a voltage is abruptly turned on, 
the voltages in the $x$ and $y$ directions temporarily vary 
in the same manner as Eqs.~(\ref{hami:jx}) and (\ref{hami:jy}), 
\textit{i.e.}, the period of the oscillation is given by the 
energy difference between two sublevels (see Appendix).  
Finally, although the electric field is given by the step function here, 
we are able to calculate the time dependence of the currents for 
an arbitrarily time-dependent electric field by following the machinery 
of the Kubo formula (See \eref{constj:flinear} below).  

\section{Long-Time Behavior of Currents \label{longtime}}

Let us study the currents after a long time.  
We show that $|\Delta\sigma_x(t)|$ and $|\Delta\sigma_y(t)|$ 
decay as $1/t$ 
using the Riemann-Lebesgue theorem\cite{RL}: 
$\lim_{t\to\infty}
\int_{\omega_a}^{\omega_b}g(\omega)\re^{\ri\omega t}\rd\omega=0$, 
where $g(\omega)$ is uniformly convergent.  

Both $\Delta\sigma_x(t)$ and $\Delta\sigma_y(t)$ are expressed as 
($\alpha=x,\,y$) 
\be
\Delta\sigma_{\alpha}(t)
=
\sum_{m\le m_0}\sum_{m'>m_0}\int_{\rm MBZ}\rd^2\vecvar{k}
\Ima\bar{g}_{mm'}^{(\alpha)}(\vecvar{k})\re^{
\ri\left(\epsilon_{m'}(\vecvar{k})-\epsilon_m(\vecvar{k})\right)t/\hbar}.
\ee
We define 
\be
\omega_{mm'}(\vecvar{k})
\equiv
\frac{\epsilon_{m'}(\vecvar{k})-\epsilon_m(\vecvar{k})}{\hbar},
\ee
\be
\omega_{0}\equiv\omega_{mm'}(k_x=0,k_y),\quad
\omega_{\pi/qa}\equiv\omega_{mm'}(k_x=\pi/qa,k_y).
\ee
Hence, 
\bea
\Delta\sigma_{\alpha}(t)
&=&
2\sum_{m\le m_0}\sum_{m'>m_0}
\int_0^{2\pi/b}\rd k_y
\int_{\omega_{0}}^{\omega_{\pi/qa}}\rd\omega_{mm'}
\left|\frac{\partial\omega_{mm'}}{\partial k_x}\right|^{-1}
\Ima\bar{g}_{mm'}^{(\alpha)}(k_x,k_y)
\re^{\ri\omega_{mm'}t}
\nonumber \\
&=&
\sum_{m\le m_0}\sum_{m'>m_0}\Ima
\int_0^{2\pi/b}\rd k_y
\int_{\omega_a}^{\omega_b}\rd\omega_{mm'}\,
g_{mm'}^{(\alpha)}(\omega_{mm'},k_y)
\re^{\ri\omega_{mm'}t},
\label{longtime:Dsigma}
\eea
where 
$\omega_a=\min\left(\omega_{0},\,\omega_{\pi/qa}\right)$ 
and 
$\omega_b=\max\left(\omega_{0},\,\omega_{\pi/qa}\right)$.  

We note that 
$\int_{\omega_a}^{\omega_b}\rd\omega\re^{\ri\omega t}=
(\re^{\ri\omega_bt}-\re^{\ri\omega_at})/\ri t$ and therefore 
this integral decays as $1/t$.  
We express $g_{mm'}^{(\alpha)}(\omega)$ as 
\be
g_{mm'}^{(\alpha)}(\omega)=
\left(u_>(\omega)+u_<(\omega)\right)+\ri
\left(v_>(\omega)+v_<(\omega)\right),
\ee
where $u_>$ and $v_>$ are positive and $u_<$ and $v_<$ are negative 
in $\omega_a<\omega<\omega_b$.  
We write the maximum and minimum of these functions as 
$u_>^{\rm max}\equiv\max|u_>(\omega)|$, 
$u_>^{\rm min}\equiv\min|u_>(\omega)|$, 
\textit{etc.}  
Then we see the integrals,
$\int_{\omega_a}^{\omega_b}\rd\omega u_>^{\rm min}\re^{\ri\omega t}$, 
$\int_{\omega_a}^{\omega_b}\rd\omega u_>^{\rm max}\re^{\ri\omega t}$, 
\textit{etc.} 
also decay as $1/t$.  
We have
\be
\left|\int_{\omega_a}^{\omega_b}\rd\omega_{mm'}
u_>^{\rm min}\re^{\ri\omega_{mm'}t}\right|
\le
\left|\int_{\omega_a}^{\omega_b}\rd\omega_{mm'}
u_>(\omega)\re^{\ri\omega_{mm'}t}\right|
\le
\left|\int_{\omega_a}^{\omega_b}\rd\omega_{mm'}
u_>^{\rm max}\re^{\ri\omega_{mm'}t}\right|,
\ee
\be
\left|\int_{\omega_a}^{\omega_b}\rd\omega_{mm'}
u_<^{\rm min}\re^{\ri\omega_{mm'}t}\right|
\le
\left|\int_{\omega_a}^{\omega_b}\rd\omega_{mm'}
u_<(\omega)\re^{\ri\omega_{mm'}t}\right|
\le
\left|\int_{\omega_a}^{\omega_b}\rd\omega_{mm'}
u_<^{\rm max}\re^{\ri\omega_{mm'}t}\right|,
\ee
\textit{etc.}  
Therefore 
\be
\left|\int_{\omega_a}^{\omega_b}\rd\omega_{mm'}
g_{mm'}^{(\alpha)}(\omega)\re^{\ri\omega_{mm'}t}\right|
\sim\frac{1}{t}.
\ee
Thus we have shown 
\be
|\Delta\sigma_{\alpha}(t)|\sim\frac{1}{t}\quad
(\alpha=x,\,y).
\ee

\begin{figure}[tb]
\begin{center}
\includegraphics[scale=1.0]{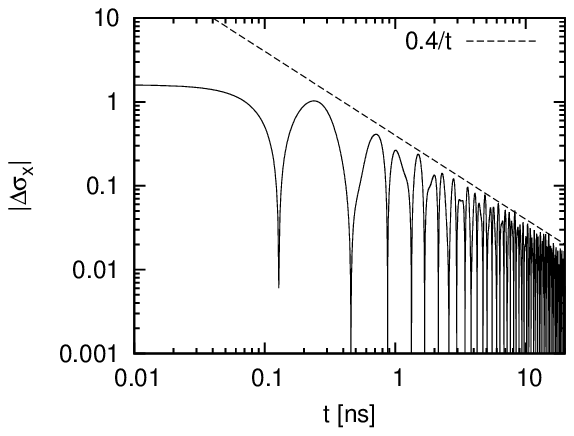}
\includegraphics[scale=1.0]{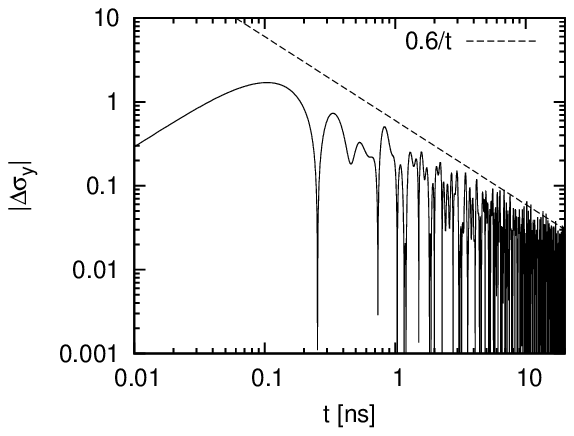}
\end{center}
\caption{
Case (i): Logarithmic plots of the long-time behavior of 
$|\Delta\sigma_x(t)|$ and $|\Delta\sigma_y(t)|$.  
We also draw the dashed lines $0.4/t$ (left) and 
$0.6/t$ (right) to see $|\Delta\sigma_x(t)|$ and $|\Delta\sigma_y(t)|$ 
decay as $1/t$.  The parameters are the same as in \fref{hami:fig02b}.
}
\label{longtime:f02log}
\end{figure}

\begin{figure}[tb]
\begin{center}
\includegraphics[scale=1.0]{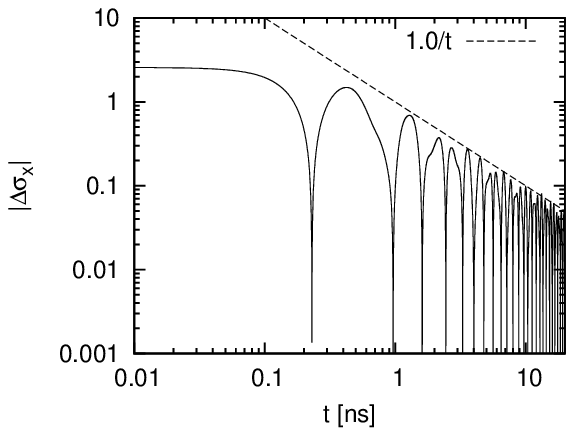}
\includegraphics[scale=1.0]{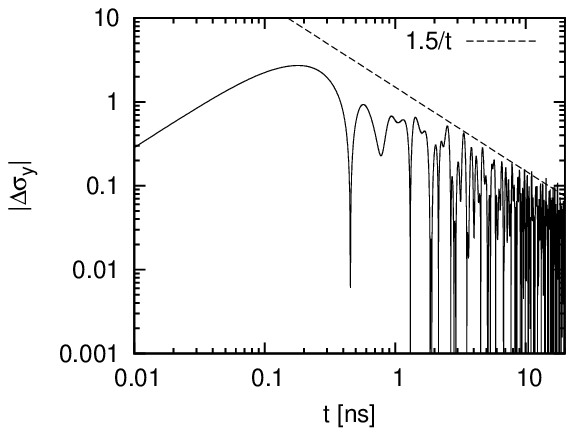}
\end{center}
\caption{
Case (ii): Logarithmic plots of the long-time behavior of 
$|\Delta\sigma_x(t)|$ and $|\Delta\sigma_y(t)|$.  
We also draw the dashed lines $1.0/t$ (left) and 
$1.5/t$ (right) to see $|\Delta\sigma_x(t)|$ and $|\Delta\sigma_y(t)|$ 
decay as $1/t$.  The parameters are the same as in \fref{hami:fig03b}.
}
\label{longtime:f03log}
\end{figure}

\begin{figure}[tb]
\begin{center}
\includegraphics[scale=1.0]{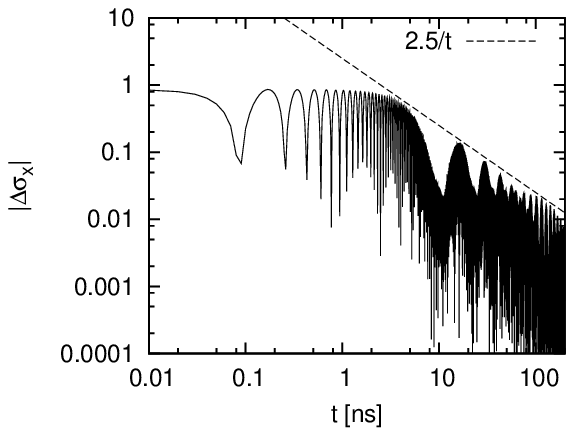}
\includegraphics[scale=1.0]{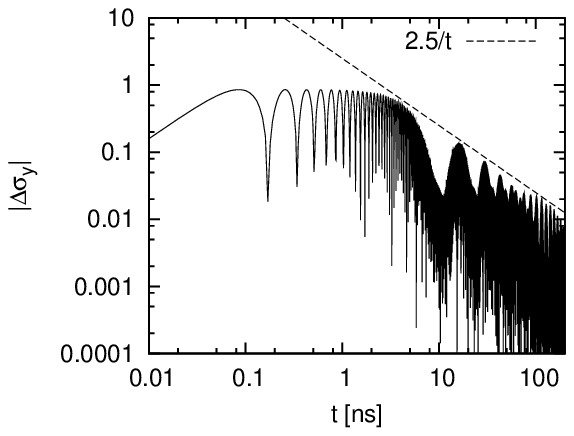}
\end{center}
\caption{
Case (ii): Logarithmic plots of the long-time behavior of 
$|\Delta\sigma_x(t)|$ and $|\Delta\sigma_y(t)|$.  
We also draw the dashed lines $2.5/t$ (left) and 
$2.5/t$ (right) to see $|\Delta\sigma_x(t)|$ and $|\Delta\sigma_y(t)|$ 
decay as $1/t$.  The parameters are the same as in \fref{hami:fig05b}.
}
\label{longtime:f05log}
\end{figure}

In Figs.~\ref{longtime:f02log}, \ref{longtime:f03log}, and 
\ref{longtime:f05log}, we show logarithmic plots of 
$|\Delta\sigma_x(t)|$ and $|\Delta\sigma_y(t)|$ in the three cases 
(i) $p/q=5/4$ and $m_0=2$ ($N_{\rm Ch}=2$),   
(ii) $p/q=7/6$ and $m_0=3$ ($N_{\rm Ch}=3$), and 
(iii) $p/q=7/6$ and $m_0=1$ ($N_{\rm Ch}=1$).  
In all cases, $|\Delta\sigma_x(t)|$ and $|\Delta\sigma_y(t)|$ 
indeed decay as $1/t$.  
Thus, the response of the system to the temporal change of the 
external field disappears in nano-second order even if 
there is no dissipative mechanism.  

Since the correction terms 
$\Delta\sigma_x(t)$ and $\Delta\sigma_y(t)$ 
decay as $1/t$, in the limit $t\to\infty$, we obtain
\bea
j_x(t\to\infty)
&=&
\frac{E_ye^2}{2\pi\hbar}N_{\rm Ch},
\nonumber \\
j_y(t\to\infty)
&=&
0.
\eea
This Hall current was first obtained by Thouless \textit{et al.}
\cite{Thouless82a}  

When the bands $\epsilon_m(\vecvar{k})$ and $\epsilon_{m'}(\vecvar{k})$ 
are nearly flat as in the case (iii), 
we can explicitly calculate the time dependence of 
$\Delta\sigma_{\alpha}(t)$.  
In this case, $\omega_b-\omega_a$ is very small and 
\eref{longtime:Dsigma} can be approximated as 
\be
\Delta\sigma_{\alpha}(t)\simeq
\sum_{m\le m_0}\sum_{m'>m_0}\Ima
\int_0^{2\pi/b}\rd k_y\,
\tilde{g}_{mm'}^{(\alpha)}(k_y)
\int_{\omega_a}^{\omega_b}\rd\omega_{mm'}\,
\re^{\ri\omega_{mm'}t},
\ee
where 
$\tilde{g}_{mm'}^{(\alpha)}(k_y)\,\left(
=g_{mm'}^{(\alpha)}(\omega_{mm'},k_y)
\right)$ 
is independent of 
$\omega_{mm'}$.  
We note that
\be
\int_{\omega_a}^{\omega_b}\rd\omega\re^{\ri\omega t}=
\frac{2}{t}\sin\left(\frac{\omega_b-\omega_a}{2}t\right)
\re^{\ri(\omega_b+\omega_a)t/2}.
\ee
This integral decays as $1/t$ and 
its amplitude has two kinds of oscillations.  
The period of one oscillation is inversely proportional to 
$\omega_b-\omega_a$ and the period of the other is inversely 
proportional to $\omega_b+\omega_a$.  
We note that the difference $\omega_b-\omega_a$ is very small and 
$\omega_a\simeq\omega_b$.  Therefore, 
the frequency of $\exp[\ri(\omega_b+\omega_a)t/2]$ is given by 
the energy difference between $\epsilon_{m}(\vecvar{k})$ and 
$\epsilon_{m'}(\vecvar{k})$, and 
the period of the beat $4\pi/(\omega_b-\omega_a)$ is very long.  
The $1/t$ decay is revealed for a time longer than the period of 
the beat.  

Thus, in nearly flat-band cases, it is easier to observe 
the $1/t$ dependence because 
$|\Delta\sigma_x(t)|$ and $|\Delta\sigma_y(t)|$ decay rather 
slowly and survive for a long time as is seen 
in \fref{longtime:f05log}.  

\section{Conclusions}

Using the Greenwood linear response theory, we studied the time 
dependence of the currents in the quantum Hall effect when 
the electric field is suddenly turned on.  
We found that both $j_x(t)$ and $j_y(t)$ oscillate because of the 
quantum fluctuation between two subbands which straddle the Fermi 
energy.  These oscillations decay as $1/t$ and 
eventually cease.  
In the limit $t\to\infty$, 
$j_x(t\to\infty)$ is given as the Chern number $N_{\rm Ch}$ 
multiplied by $e^2/h$ as Thouless \textit{et al.}\cite{Thouless82a} 
obtained.  
As is discussed in Appendix, 
the electric fields oscillate in time in the same 
way as $j_x(t)$ and $j_y(t)$ when, in reverse, the current is 
applied abruptly at $t=0$.  

We showed that the ratio of the Hall current and 
the suddenly applied dc field 
is decomposed into the sum of a constant term and a time-dependent term 
(\eref{hami:jx}).  
The constant term is the conductivity for the dc field applied 
for infinite time 
and given by the Chern number $N_{\rm Ch}$.  Thus, the time-dependent 
term $\Delta\sigma_x(t)$ expresses a correction to the Chern number term.  
In other words, $\Delta\sigma_x(t)$ can be regarded as the fluctuation 
around the Chern number.  
It will be remarkable to observe the fluctuation experimentally, since 
the quantization to the Chern number is normally regarded as 
very rigid.  
This fluctuation, which stems from transitions 
between different subbands, decays as $1/t$.  

Thus, the response of the system to the temporal change of the 
external field decays as $1/t$ even if there is no dissipative mechanism.  
The amplitude of the decay 
gets large if the bands that give large contribution to 
$\Delta\sigma_x(t)$ are nearly flat.  In this case, the $1/t$-decay 
survives for a long time.  
In a quantum Hall system on a semiconductor heterojunction, 
this power-law decay of the order of nano-second might be difficult to 
observe experimentally because relaxation time due to 
impurity scattering, \textit{etc.} is of pico-second order.
\cite{Coleridge91}  
Cold atomic systems under an artificial magnetic field may 
overcome these difficulties.  
In experiments of a Rubidium cold atomic gas trapped by 
a rotating optical lattice, 
the time scale of the power-law decay is of the order of millisecond 
for $a=b\sim 1\mu{\rm m}$, $U_0\sim 0.1{\rm neV}$, and large $\Omega$ 
so that $p/q\sim 1$.

\section*{Acknowledgment}
One of the authors (M. M.) thanks Dr.~Mikito Koshino 
for indispensable discussions on the numerical calculation 
of the Hall current.  
M. M. is also grateful to Prof.~Piet W. Brouwer for discussion on 
linear response theory and to Dr.~Tomio Y. Petrosky for letting him know 
the Riemann-Lebesgue theorem.  
This study is supported by Grant-in-Aid for Scientific Research 
(No.~17340115 and No.~19740241) 
from the Ministry of Education, Culture, Sports, 
Science and Technology as well as by Core Research for 
Evolutional Science and Technology (CREST) of Japan Science and 
Technology Agency.  
The computation in this work was carried out partly on 
the facilities of the Supercomputer Center, Institute 
for Solid State Physics, the University of Tokyo.

\appendix
\section{Measuring Electric Fields under an Applied Current\label{constj}}

Here, we calculate the voltage for the applied dc current 
that is switched on abruptly.  
This situation matches current-controlled experiments.  
(For theoretical reasons, in the main body of the paper, we calculate 
the current under the applied voltage.)  
We find that the voltage also oscillates.  
Both temporal oscillations of the current and voltage are caused 
by the quantum fluctuation between two subbands.  

We apply the current suddenly at $t=0$ in the 
$y$ direction, $j_x(t)=0,\;j_y(t)=J_y\theta(t)$, and 
obtain $E_x(t)$ and $E_y(t)$.  This may be closer to the 
experimental situation.  

Since we assume linear response, we have ($\alpha=x,\,y$)
\be
j_{\alpha}(t)=
\sum_{\beta=x,y}\int_{-\infty}^{\infty}\rd t'\, 
\sigma_{\alpha\beta}(t-t')E_{\beta}(t').
\ee
By Fourier transform, we obtain 
\be
\tilde{j}_{\alpha}(\omega)=
\sum_{\beta=x,y}
\tilde{\sigma}_{\alpha\beta}(\omega+\ri\eta)\tilde{E}_{\beta}(\omega),
\label{constj:flinear}
\ee
where 
$\tilde{j}_{\alpha}(\omega)=
\int j_{\alpha}(t)\re^{\ri\omega t}\rd t$, 
\textit{etc.}  We put an infinitesimally small $\eta>0$ 
to ensure the causality: $\sigma_{\alpha\beta}(t)=0$ for $t<0$.  
We define the resistivity $\tilde{\rho}_{\alpha\beta}$ as 
\be
\tilde{E}_{\alpha}(\omega)=
\sum_{\beta=x,y}
\tilde{\rho}_{\alpha\beta}(\omega+\ri\eta)\tilde{j}_{\beta}(\omega),
\label{constj:flinearrho}
\ee
where 
\bea
\tilde{\rho}_{xy}(\omega+\ri\eta)
&=&
\frac{-\tilde{\sigma}_{xy}(\omega+\ri\eta)}
{\tilde{\sigma}^{2}_{xy}(\omega+\ri\eta)+
\tilde{\sigma}^{2}_{yy}(\omega+\ri\eta)},
\nonumber \\
\tilde{\rho}_{yy}(\omega+\ri\eta)
&=&
\frac{\tilde{\sigma}_{yy}(\omega+\ri\eta)}
{\tilde{\sigma}^{2}_{xy}(\omega+\ri\eta)+
\tilde{\sigma}^{2}_{yy}(\omega+\ri\eta)}.
\eea
Therefore we obtain electric fields as 
\bea
E_x(t)
&=&
\frac{1}{2\pi}\int\rd\omega
\frac{-\tilde{\sigma}_{xy}(\omega+\ri\eta)}
{\tilde{\sigma}^{2}_{xy}(\omega+\ri\eta)+
\tilde{\sigma}^{2}_{yy}(\omega+\ri\eta)}
\tilde{j}_y(\omega)\re^{-\ri\omega t},
\label{constj:Expre}
\\
E_y(t)
&=&
\frac{1}{2\pi}\int\rd\omega
\frac{\tilde{\sigma}_{yy}(\omega+\ri\eta)}
{\tilde{\sigma}^{2}_{xy}(\omega+\ri\eta)+
\tilde{\sigma}^{2}_{yy}(\omega+\ri\eta)}
\tilde{j}_y(\omega)\re^{-\ri\omega t},
\label{constj:Eypre}
\eea
where
\be
\tilde{j}_y(\omega)=\frac{\ri J_y}{\omega+\ri\eta}.
\ee

We obtain the conductivities $\tilde{\sigma}_{xy}(\omega)$ and 
$\tilde{\sigma}_{yy}(\omega)$ with the help of the calculation 
in \S\ref{hami}.  We first note that 
\be
\tilde{E}_y(\omega)=
\int\rd t E_y\theta(t)\re^{\ri(\omega+\ri\eta)t}=
\frac{\ri E_y}{\omega+\ri\eta}.
\ee
Using Eqs.~(\ref{hami:jx}) and (\ref{hami:jy}), we have 
\bea
\tilde{\sigma}_{xy}(\omega+\ri\eta)
&=&
\frac{\tilde{j}_x(\omega)}{\tilde{E}_y(\omega)}
\nonumber \\
&=&
\frac{e^2}{2\pi\ri\hbar}(\omega+\ri\eta)\Biggl[
\frac{\ri N_{\rm Ch}}{\omega+\ri\eta}+
\left(\sum_{m\le m_0}\sum_{m'> m_0}-
\sum_{m> m_0}\sum_{m'\le m_0}\right)
\nonumber \\
&\times&
\int_{\rm MBZ}\frac{\rd^2\vecvar{k}}{2\pi}
\biggl\langle
\frac{\partial u_{0m}(\vecvar{k})}{\partial k_y}
\biggm|
{u_{0m'}(\vecvar{k})}
\biggr\rangle
\biggl\langle
u_{0m'}(\vecvar{k})
\biggm|
\frac{\partial u_{0m}(\vecvar{k})}{\partial k_x}
\biggr\rangle
\nonumber \\
&\times&
\frac{1}{
\omega+\ri\eta+
\left(\epsilon_{m'}(\vecvar{k})-\epsilon_{m}(\vecvar{k})\right)/\hbar
}
\Biggr]
\nonumber \\
\label{constj:fsxy}
\\
\tilde{\sigma}_{yy}(\omega+\ri\eta)
&=&
\frac{\tilde{j}_y(\omega)}{\tilde{E}_y(\omega)}
\nonumber \\
&=&
\frac{e^2}{2\pi\ri\hbar}(\omega+\ri\eta)
\left(\sum_{m\le m_0}\sum_{m'> m_0}-
\sum_{m> m_0}\sum_{m'\le m_0}\right)
\nonumber \\
&\times&
\int_{\rm MBZ}
\frac{\rd^2\vecvar{k}}{2\pi}
\left|\biggl\langle
\frac{\partial u_{0m}(\vecvar{k})}{\partial k_y}
\biggm|
u_{0m'}(\vecvar{k})
\biggr\rangle\right|^2
\nonumber \\
&\times&
\frac{1}{
\omega+\ri\eta+
\left(\epsilon_{m'}(\vecvar{k})-\epsilon_{m}(\vecvar{k})\right)/\hbar
}.
\label{constj:fsyy}
\eea
By plugging Eqs.~(\ref{constj:fsxy}) and (\ref{constj:fsyy}) into 
Eqs.~(\ref{constj:Expre}) and (\ref{constj:Eypre}), we obtain 
the electric fields: 
\bea
E_x(t)
&=&
\frac{2\pi\hbar}{e^2}J_y
\int\frac{\rd^2\vecvar{k}}{\pi}\sum_{m\le m_0}\sum_{m'>m_0}
\tilde{S}\left(\frac{\epsilon_{m'}(\vecvar{k})-\epsilon_{m}(\vecvar{k})}
{\hbar}\right)
\nonumber \\
&\times&
\Ima
\biggl\langle
\frac{\partial u_{0m}(\vecvar{k})}{\partial k_y}
\biggm|
u_{0m'}(\vecvar{k})
\biggr\rangle
\biggl\langle
u_{0m'}(\vecvar{k})
\biggm|
\frac{\partial u_{0m}(\vecvar{k})}{\partial k_x}
\biggr\rangle
\re^{\ri\left(\epsilon_{m'}(\vecvar{k})-
\epsilon_{m}(\vecvar{k})\right)t/\hbar},
\label{constj:Ex}
\\
E_y(t)
&=&
\frac{2\pi\hbar}{e^2}J_y
\int\frac{\rd^2\vecvar{k}}{\pi}\sum_{m\le m_0}\sum_{m'>m_0}
\tilde{S}\left(\frac{\epsilon_{m'}(\vecvar{k})-\epsilon_{m}(\vecvar{k})}
{\hbar}\right)
\nonumber \\
&\times&
\left|
\biggl\langle
\frac{\partial u_{0m}(\vecvar{k})}{\partial k_y}
\biggm|
u_{0m'}(\vecvar{k})
\biggr\rangle
\right|^2
\sin\left[
\left(\epsilon_{m'}(\vecvar{k})-
\epsilon_{m}(\vecvar{k})\right)t/\hbar)\right],
\label{constj:Ey}
\eea
where
\be
\tilde{S}(\omega)^{-1}=
-\left(\frac{2\pi\hbar}{e^2}\right)^2\left[
\tilde{\sigma}_{xy}^2(\omega)+\tilde{\sigma}_{yy}^2(\omega)
\right].
\ee
Note that $E_{x}(t)$ and $E_{y}(t)$ have the same 
time dependence as $j_{x}(t)$ and $j_{y}(t)$ in \S\ref{hami}; 
the period of the oscillation is dominantly given by the energy difference 
between two subbands which straddle the Fermi energy.

\end{document}